\documentclass[preprint]{revtex4-1} 
\usepackage{graphicx}  % needed for figures
\usepackage{dcolumn}   % needed for some tables
\usepackage{bm}        % for math
\usepackage{amssymb}   % for math
\usepackage{epstopdf} %converts eps to pdf
\usepackage{hyperref}
\usepackage{amsmath}
\usepackage{float}

\begin{document}
%\setboolean{@twoside}{false}
%\hspace{3cm}\textbf{Manuscript Classification:} Physical Sciences/Physics
%\title{Hidden angular momentum in interacting many-body systems}
%\title{Angular momentum in interacting many-body systems hides in phantom vortices}
\title{Phantom vortices: hidden angular momentum in ultracold dilute Bose-Einstein condensates}
\author{Storm E. Weiner$^{\dagger}$\footnote[0]{$^\dagger$
Corresponding Author: StormWeiner@Berkeley.edu}
$^{1}$\footnote[0]{$^1$Department of Chemistry, University of California at Berkeley, CA, USA}}
\author{Marios C. Tsatsos $^2$\footnote[0]{$^2$Instituto de F\'isica de S\~ao Carlos, Universidade de S\~ao Paulo, S\~ao Carlos, S\~ao Paulo, Brazil}}                          
\author{Lorenz S. Cederbaum$^3$\footnote[0]{$^3$Theoretische Chemie, Physikalisch--Chemisches Institut, Universit\"at Heidelberg, Germany}}
%,Im Neuenheimer Feld 229, D-69120 Heidelberg, Germany}}
\author{Axel U. J. Lode$^4$\footnote[0]{$^4$Department of Physics, University of Basel, Switzerland}}
%, Klingelbergstrasse 82, CH-4056 Basel, Switzerland}}

\date{\today}

\begin{abstract}
 Vortices are essential to angular momentum in quantum systems such as 
 ultracold atomic gases.
 The existence of quantized vorticity in bosonic systems stimulated the development of the Gross-Pitaevskii mean-field approximation.
 However, the true dynamics of angular momentum in finite, interacting many-body systems like trapped Bose-Einstein condensates is enriched by the emergence of quantum correlations whose description demands more elaborate methods.
 Herein we theoretically investigate the full many-body dynamics of the acquisition of angular momentum by a gas of ultracold bosons in two dimensions using a standard rotation procedure. 
 We demonstrate the existence of a novel mode of quantized vorticity, which we term the \textit{phantom vortex} that, contrary to the conventional mean-field vortex, can be detected as a topological 
 defect of spatial coherence, but \textit{not} of the density. 
 We describe previously unknown many-body mechanisms of vortex nucleation and show that angular momentum is hidden in phantom vortex modes which so far seem to have evaded experimental detection.\\
 \end{abstract}

%Bose-Einstein condensation
%dynamic properties, 03.75.Kk
%entanglement and decoherence, 03.75.Gg
%multicomponent and spinor condensates, 03.75.Mn
%quantum optics, 42.50.Gy
%solitons, 03.75.Lm
%static properties, 03.75.Hh
%tunneling, 03.75.Lm
%vortices in, 03.75.Lm
%Topological excitations (Bose-Einstein condensation), 03.75.Lm

%05.30.Jp Boson systems

\maketitle

\clearpage
\renewcommand{\thefootnote}{\fnsymbol{footnote}}
\section*{Introduction}
Quantized vortices are perhaps the most interesting way angular momentum is known to manifest in quantum many-body systems~\cite{Pethick,Ketterle2001}. They appear in a variety of experiments with systems including %superfluid Helium~\cite{Donnelly}, type-II superconductors~\cite{Abrikosov1957}
atomic Bose-Einstein condensates (BECs)~\cite{Vortex_Exp,Vortex_Exp2,Ketterle2001,Kali}, and exciton-polariton condensates~\cite{Baas2008}. Typically, a quantum vortex is characterized by a density node and a phase discontinuity. Other known signatures of angular momentum, such as center-of-mass rotation, surface waves, and quadrupole modes are also features of the density profile~\cite{Dalibard2000,Pethick}. However, it is well known that there exist aspects of many-body dynamics, like fragmentation~\cite{James1982,Spekkens1999}, that may not be visible in the density~\cite{glauber,Kaspar2008}. Coreless vortices, i.e., vortices invisible in the density, are known to exist in mean-field spinor condensates~\cite{Lovegrove2014,Kasamatsu2003} %, including superfluid Helium-3 ~\cite{}, 
and have been recently found in a multicomponent condensate ~\cite{Liu2013}. Since the core of a vortex in one (spin-)component is filled with the other (spin-)component, coreless vortices are not visible in the density. However, they have not yet been seen in single-component condensates with no spin.
%[MARIOS
Note that experiments studying vortex nucleation in rotating BECs have thus far observed a critical frequency below which 
%focused on the detection of quantized vortices beyond a critical value for the rotation velocity. 
%--]
%Below this critical velocity 
no vortices are detected, see for instance~\cite{Ketterle2001,Hodby,Feder}. As we shall show in the following, however, vortices without cores are present in single-component BECs even below the critical rotation velocity.
%[MARIOS 
%--]
These coreless vortices, to date, have evaded experimental detection as they cannot be seen from the density.

Traditionally, the dynamics of BECs are aptly treated using the time-dependent Gross-Pitaevskii (GP) equation~\cite{Dalfovo1999} which assumes that the many-body wavefunction is coherent and condensed for all time. Naturally, this assumption neglects the existence of quantum correlations and precludes the possibility of fragmentation, i.e.~macroscopic occupation of more than one single particle state~\cite{Onsager1956}. 
%[Storm
In other computational investigations, the rotation frequency was modulated at fixed anisotropy which also leads to instabilities and formation of vortex lattices, within the GP theory~\cite{Tsubota2002,Suominen2003,Parker2005}, the Bogoliubov approach~\cite{Lobo2004} and beyond~\cite{Wright2008}.
%]
More recently, much effort has been devoted to exploring the role of fragmentation in stationary states of bosonic systems~\cite{Spekkens1999,Cederbaum2003,Streltsov2006,Mueller2006,Lewenstein2009,Fischer2009,Fischer2014} including those with spin degrees of freedom~\cite{Qi2013,Song2014} using methods such as best mean-field theory or general variational many-body approaches. In particular, it has been shown that there exist energy eigenstates with definite angular momentum that are fragmented for repulsive interactions~\cite{Mueller2006} as well as in two dimensions for attractive 
interaction~\cite{Marios2014}. Furthermore, in the study of adiabatic vortex nucleation in few-body systems~\cite{Lewenstein2009}, it was shown that fragmentation is unavoidable. Generally, nontrivial dynamics involve many eigenstates, each of which may be fragmented. Hence, it stands to reason that a true many-body method is necessary to describe the dynamics of a two-dimensional condensate as it acquires angular momentum.

An appropriate many-body method is the multiconfigurational time-dependent Hartree for bosons (MCTDHB)~\cite{MCTDHB} method, which has been shown to accurately capture the transition from coherence to fragmentation with great success in a variety of systems~\cite{Alexej2007,Axel2012,streltsova2014}. MCTDHB provides highly accurate results even for problems with time-dependent traps and time-dependent interparticle interactions \cite{Lode2012,AxelThesis}.

Here we investigate the dynamics of interacting bosons in two dimensions by numerically solving the time-dependent many-body Schr\"odinger equation for a standard rotation scenario using the recursive implementation of MCTDHB in the MCTDH-X software package~\cite{RMCTDHB}. Our work is among the first that explore the many-body dynamics of this type of system beyond mean-field theory~\cite{MariosJLTP}.
The system that we describe here is similar to that simulated in Ref.~\cite{Suominen2003} using the time-dependent GP equation, and analyzed experimentally in Ref.~\cite{Madison2000}. There, regions in the parameter space were established for which vortices are observed. Even though the GP treatment for the parameters chosen herein predicts no vortex nucleation (Supplementary Information), the many-body analysis exhibits rich vortex dynamics that are \textit{invisible in the density}. Our results demonstrate that angular momentum in fragmented condensates manifests itself in a new type of coreless vortex, which we name the \textit{phantom vortex} due to its elusive nature. It is important to stress that we found phantom vortices for a wide range of particle numbers, interaction strength, interaction range, trap anisotropy, and rotation frequencies, see (Supplementary Information). An analysis of the energy eigenstates of a six-boson system in the corotating frame has shown signatures of 
phantom vortices (see Fig.~3 in 
Ref.~\cite{Lewenstein2009}). Yet, their importance in the dynamical acquisition of angular momentum by many-body systems, even in the parameter regime where no vortices nucleate in the density, has been overlooked. 

\section*{System Description}

We begin our analysis, \textit{in silico}, by first computing the many-body ground state of interacting bosons in an isotropic harmonic trap.  Subsequently, we gently transfer angular momentum into the system by elliptically deforming the harmonic trap while rotating the axis of anisotropy at a fixed frequency. The anisotropy parameter is varied in a piecewise-linear fashion; first, it is ramped from zero to a maximum value, then held constant over a fixed time interval, and finally ramped back to zero.  The trap is then kept isotropic for the remainder of the simulation. See Fig.~\ref{fig:Lz}a and Supplementary Information for details.

%[MARIOS
The parameters chosen here correspond to rotating, weakly interacting, zero-temperature atoms, confined by a pancake-like trap. The particle density and interactions are similar to the small (i.e. $N<500$) $^{87}$Rb condensate reported in \cite{Schmied2016}. The subcritical rotation frequency chosen herein is not fast enough to nucleate quantized vortices in the density. A GP ($M=1$) simulation with the same parameters only absorbs $0.5\hbar$ of angular momentum per particle (see Fig~1.d).
%]

The dynamics of an interacting gas of $N$ bosons are governed by the time-dependent many-body Schr\"odinger equation
\begin{equation}
 i \partial_t \vert \Psi\rangle = \hat{\mathcal{H}} \vert \Psi\rangle .
 \label{eq:shro}
\end{equation}
We solve Eq.~\ref{eq:shro} using the MCTDHB ansatz, written in second quantization as
\begin{equation}
 \vert \Psi \rangle = \sum_{n_1+...+n_M=N} C_{n_1,...,n_M}(t) \vert n_1,...,n_M;t\rangle. \label{ansatz}
\end{equation}
That is, the bosons dynamically populate all configurations of $N$ particles in $M$ time-dependent variationally optimized single-particle states, called orbitals (Supplementary Information). The case where $M=4$ and $N=100$ is analyzed in the main text of this work whereas other $N,\ M$ are discussed in the Supplementary Information.
In our setting the particle density at rest ($t=0$) is close to a Gaussian-like density distribution because the bosons are weakly interacting. The temperature is absolute zero throughout the dynamics and the external frequency of rotation of the mild anisotropy is at $78\%$ of the trapping frequency.

The many-body Hamiltonian in dimensionless units reads:
\footnotetext{To convert to dimensionful units, the Hamiltonian is multiplied by $\frac{\hbar^2}{m L^2}$. With the mass $m=1.44316\times10^{-25}$Kg corresponding to $Rb^{87}$, and length scale $L=0.750\mu$m, we get a transverse trapping frequency of $\omega_0=1.30 \times 10^3 s^{-1}$ and a time scale of $4.84 $ms. With this choice of scale, $t=500$ corresponds to $2.42 s$, and the linear extent of the simulation is $12 \mu$m. The interaction parameter $\lambda_0$ is related to the scattering length $a_s$ and transverse confinement $l_z=\sqrt{\frac{\hbar}{m\omega_z}}$ by $\lambda_0=2\sqrt{2\pi} \frac{a_s}{l_z}$. For $Rb^{87}$, $a_s=90.4a_0$. Using this scattering length and $\lambda_0(N-1)=17.1$ gives $\omega_z=37.9 $kHz for $N=100$. So the transverse aspect ratio is $\frac{\omega_0}{\omega_z}=3.43 \times 10^{-2}$ and the maximum energy per particle, $E/\hbar N=1.41 $kHz is merely $3.72\times 10^{-2} \omega_z$. Thus, we can truly treat the dynamics as two-dimensional.}
  
\begin{equation}
 \hat{\mathcal{H}}(\mathbf{r}_1,...,\mathbf{r}_N;t)=\sum_{i=1}^{N} \hat{h}_i (\mathbf{r}_i;t)+ \lambda_0 \sum_{i<j}^N\hat{W}(\mathbf{r}_i-\mathbf{r}_j).
\end{equation}
We use as a two-body potential $\hat{W}$ a normalized Gaussian of width $\sigma=0.25$ and interaction strength $\lambda_0 (N-1)=17.1$. This choice of a short-range interaction is motivated by the similarity of the physics for zero- and short-ranged interaction potentials in Ref.~\cite{rosti2013}. See Supplemental Material for a direct comparison. The one-body Hamiltonian $\hat{h}_i(\mathbf{r}_i;t)$ is the sum of the kinetic energy $\hat{T}_i=-\frac{1}{2}\partial_{\mathbf{r}_i}^2$ and the trapping potential,
\begin{equation}
\hat{V}(\mathbf{r},t)= \frac{1}{2} \left(x(t)^2+y(t)^2\right)+\frac{1}{2} \eta(t) \left(x(t)^2-y(t)^2\right),
\end{equation}
where 
\begin{equation}
 \left(
\begin{array}{c}
x(t)\\
y(t)\\
\end{array}
\right) =
\left(
\begin{array}{c c}
cos(\omega t) & sin(\omega t)\\
-sin(\omega t) & cos(\omega t)\\
\end{array}
\right)
\left(
\begin{array}{c}
x\\
y\\
\end{array}
\right)
\end{equation} 
and $\omega=0.78$. The anisotropy parameter varies in time so that it is ramped up from $0$ to $\eta_{max}=0.1$ over $t_r=80$, held fixed at $\eta_{max}$ for $t_f=220$, ramped down to zero over $t_r$, and then held fixed at $0$ for the rest of simulation until $t=500$ as shown in Fig.~\ref{fig:Lz}a. The resulting healing length, $\xi=\frac{1}{\sqrt{2\lambda_0 (N-1)}}=0.17$ is comparable to the oscillator length (1 in dimensionless units) so the particles are indeed weakly interacting.

The primary object of our analysis is the one-body reduced density matrix (RDM), which is defined as the partial trace of the $N$-body density~\cite{Kaspar2008,glauber}:
\begin{equation}
 \rho^{(1)}(\mathbf{r}_1\vert\mathbf{r}'_1;t)  = N \int \Psi(\mathbf{r}_1,\mathbf{r}_{2},...,\mathbf{r}_N;t) 
 \times  \Psi^*(\mathbf{r}'_1,\mathbf{r}_2,...,\mathbf{r}_N;t)d\mathbf{r}_{2}...\mathbf{r}_N,
\label{eq:RDM}
\end{equation}
with $\Psi$ normalized to $1$.  The trace operation can be understood as eliminating knowledge of the many-body effects to obtain a single-particle operator, the RDM. The diagonal, $\mathbf{r}=\mathbf{r'}$, of the RDM is the particle density, $\rho(\mathbf{r};t)=\rho^{(1)}(\mathbf{r}\vert\mathbf{r};t)$, or simply the density. 
The RDM is written in its eigenbasis as
\begin{equation}
 \rho^{(1)}(\textbf{r}\vert \mathbf{r}';t)=\sum_{i=1}^M \rho^{(NO)}_i(t) \phi_i^{*}(\mathbf{r}';t) \phi_i(\textbf{r};t)
 \end{equation}
The eigenvalues, $\rho^{(NO)}_i(t)$, are ordered in decreasing magnitude and called \textit{natural occupations} and the eigenfunctions, $\phi_i(\mathbf{r};t)$, are called the \textit{natural orbitals} or synonymously, fragments. Note that the density, $\rho(\mathbf{r};t)=\rho^{(1)}(\mathbf{r}\vert \mathbf{r};t)$, is equal to the sum of the squared amplitudes of the natural orbitals, weighted with their natural occupations.

%This treatment bears some resemblance to the two-fluid model of Helium below its lambda point, where the total density has a superfluid component and thermal component interacting through a mutual viscosity.  The MCTDHB treatment can likewise be thought of as an M-fluid model, where there are M interpenetrating superfluid components coupled through the MCTDHB equations of motion~\cite{MCTDHB}, but no thermal component. In this treatment, there are hence $M$ order parameters, one for each fragment.

A state is considered coherent when only one natural orbital has significant occupation. In this case, the full information of the many-body wavefunction is contained in a single one-particle state and thus, the Gross-Pitaevskii mean-field approximation is valid. Otherwise, multiple orbitals have macroscopic occupation and the state is said to be fragmented because the density is a sum of multiple single-particle functions. We emphasize that, from the view of the RDM, the MCTDHB approach generalizes the mean-field approximation by allowing a dynamical transition from coherence to fragmentation. This means that, contrary to the mean-field case, the natural occupations are allowed to have non-integer values and vary in time. 

%[Axel
Importantly, MCTDHB boils down to the GP mean-field for the case of $M=1$, because then the ansatz for the method, Eq.~\eqref{ansatz}, becomes identical to the ansatz of the GP mean-field. This allows us to straightforwardly compare the predictions of the two approaches.
%]Axel

Another way of describing coherence is through the first-order correlation function, $g^{(1)}(\mathbf{r}\vert\mathbf{r}';t)$, defined in terms of the RDM as
\begin{equation}
 g^{(1)}(\mathbf{r}\vert\mathbf{r}';t)=\frac{\rho^{(1)}(\mathbf{r}\vert\mathbf{r}';t)}{\sqrt{\rho^{(1)}(\mathbf{r}\vert\mathbf{r};t) \rho^{(1)}(\mathbf{r}'\vert\mathbf{r}';t)}} .
\end{equation}
A coherent state at time $\tau$ must satisfy $\vert g^{(1)}(\mathbf{r}\vert\mathbf{r}';\tau)\vert=1$, while any other value indicates fragmentation~\cite{glauber}. Intuitively, $\vert g^{(1)}\vert$ measures how well the many-body wavefunction is described by a single particle state. 

The off-diagonal, $\mathbf{r} \neq \mathbf{r}'$, terms in $g^{(1)}$ and the orbitals $\phi_i$ are complex-valued.  Thus it is instructive to plot their magnitudes and phases separately. We hence define the phase of $g^{(1)}(\mathbf{r}\vert\mathbf{r}';t)$ to be 
\begin{equation}
S_{g}(\mathbf{r}\vert\mathbf{r}';t)\equiv \text{arg}\left[g^{(1)}(\mathbf{r}\vert\mathbf{r}';t)\right],
\end{equation}
and the orbital phases to be
\begin{equation}
 S_i(\mathbf{r};t)=\text{arg}\left[\phi_i(\mathbf{r};t) \right].
\end{equation}
For the sake of completeness, we likewise define the phase of the many-body wavefunction, $\Psi$, as
\begin{equation}
 S_{MB}(\mathbf{r}_1,...,\mathbf{r}_N;t)=\text{arg}\left[\Psi(\mathbf{r}_1,...,\mathbf{r}_N;t)\right].
\end{equation}

%For visualization purposes, we define $S_g(\mathbf{r}\vert\mathbf{r}';t)$ as the phase of the one-body correlation function.

\section*{Simulation Results}
\begin{figure}[H]
 \includegraphics[width=1\textwidth]{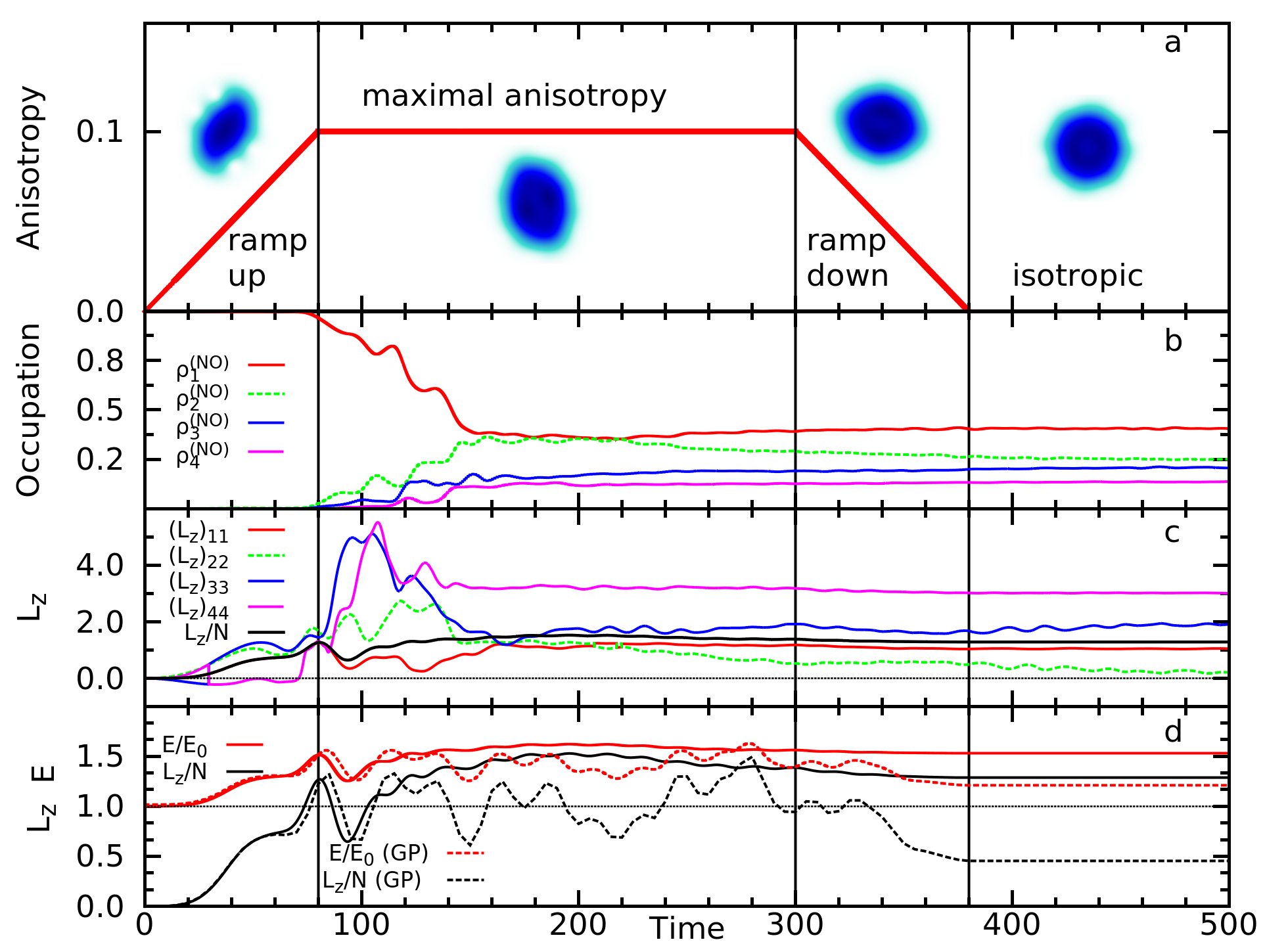}
 \caption{
 [COLOR]
 (a) Sketch of the ramping procedure of the anisotropy parameter $\eta(t)$ with plots of the density $\rho(\mathbf{r})$ at representative times in each part.
 (b) The onset of fragmentation occurs around t=80, which corresponds to the time of maximum anisotropy. By the end of the time of maximal anisotropy, the state is completely fragmented, $\rho_1^{(NO)}\approx 40\%$.
 (c) Orbital angular momenta reach their maximum values early in the period of maximum anisotropy ($t=80-150$), but evolve to their equilibrium values before the anisotropy is removed.  Discontinuities in $(L_z)_{ii}$ arise from occupation reordering [see panel (b)]. Fluctuations in $(L_z)_{22}$ and $(L_z)_{33}$ from t=250 to t=500 are strongly correlated due to vorticity transfer between $\phi_2$ and $\phi_3$.
 (d)  Comparison of energy and angular momentum curves for $M=4$ (solid lines) and $M=1$ (GP, dotted lines). Both quantities agree between the two cases until around $t=80$ when fragmentation becomes significant and the GP ansatz breaks down.   Energy and angular momentum oscillate about their maximal values by t=200, while the anisotropy is still maximal. The angular momentum stabilizes at $L_z\approx1.25$ per particle ($M=4$) and $L_z< 0.5$ per particle ($M=1$). The strong correlation between energy and $L_z$ indicates that the perturbation strictly excites angular momentum modes in both simulations. 
 To guide the eye, on the bottom plot is marked with a horizontal line at 1 on the vertical axis. All quantities shown are dimensionless.}
 \label{fig:Lz}
\end{figure}

\begin{figure}[H]
 \includegraphics[width=1\textwidth]{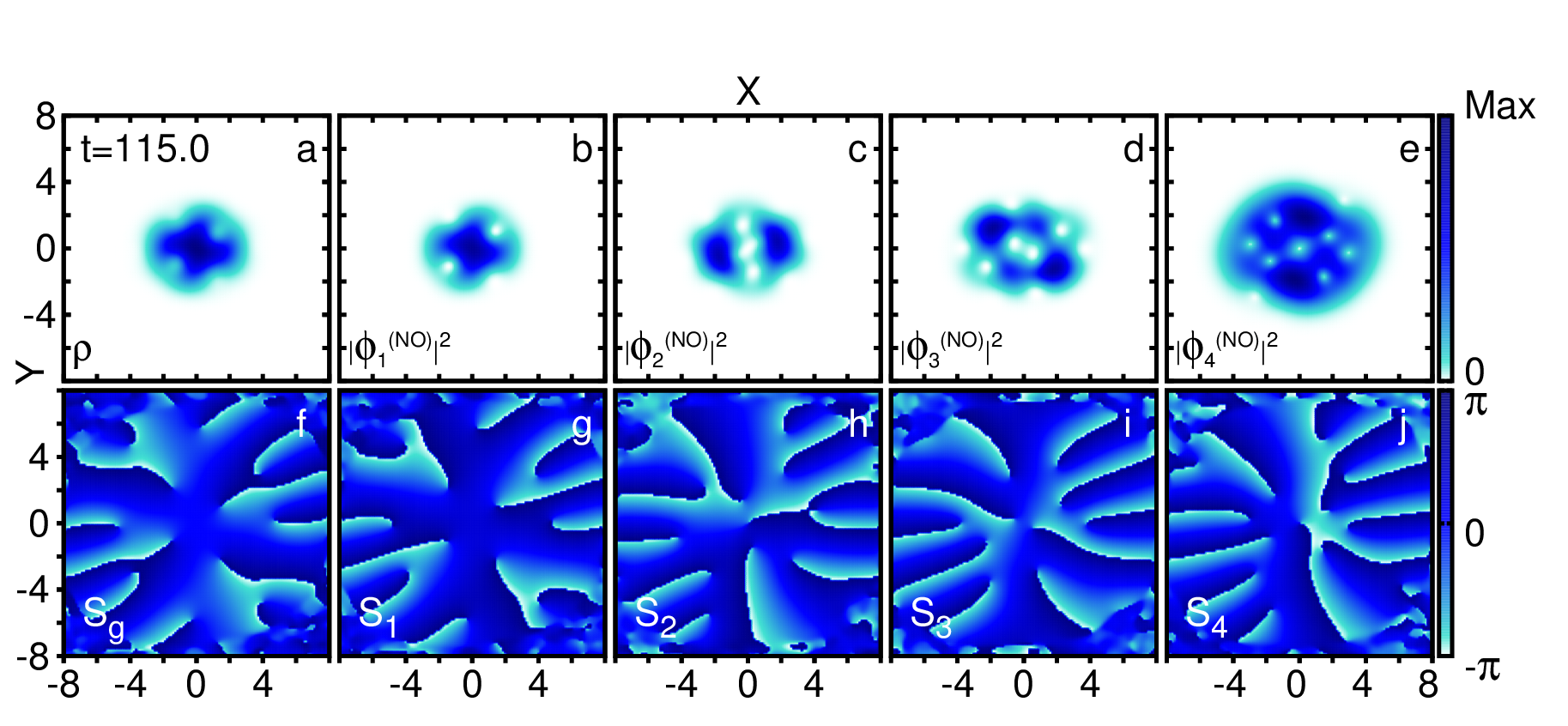}
 \caption{[COLOR]
 (a) The density $\rho(\mathbf{r})$ shows some density minima due to phantom vortices in $\phi_1$, but no true density node.
 (b)-(e) The natural orbital densities, $\vert \phi_i (\mathbf{r})\vert ^2$ ($i=1,2,3,4$), are plotted.  Many phantom vortices are present in each orbital.
 (f) The phase, $S_g(\mathbf{r}\vert0)$, of $g^{(1)}(\mathbf{r}\vert0)$ is plotted. Note, $S_{g}$ is not the many-body phase, $S_{\text{MB}}(\mathbf{r}_1,...,\mathbf{r}_N;t)$, which is too complicated to visualize. Since $\phi_1$ carries most of the particles at this time, $S_g$ bears strong resemblance to $S_1$.
 (g)-(j) The phases, $S_i$, of the natural orbitals $\phi_i$ ($i=1,2,3,4$) are plotted. Each $2\pi$ phase discontinuity marks a phantom vortex core.
The central phantom vortex in $\phi_2$ (c),(h) was mutated from its initial angular node, and persisted for the duration of simulation.  The two central phantom vortices in $\phi_3$ (d),(i) were mutated from its initial angular nodes and are transient. The three centermost phantom vortices in $\phi_4$ (e),(j) were mutated from an ``I'' shaped node and persisted for the length of the simulation.  All other phantom vortices nucleated at the edge of their orbital density and are transient. 
 All panels are plotted at $t=115.0$, when $\rho^{(NO)}_1=82.0\%$, $\rho^{(NO)}_2=11.8\%$, $\rho^{(NO)}_3=4.1\%$, and $\rho^{(NO)}_4=2.1\%$. See complementary Video S1 (Supplementary Information). All quantities shown are dimensionless.}
 \label{fig:115}
\end{figure}

\begin{figure}[H]
  \includegraphics[width=1\textwidth]{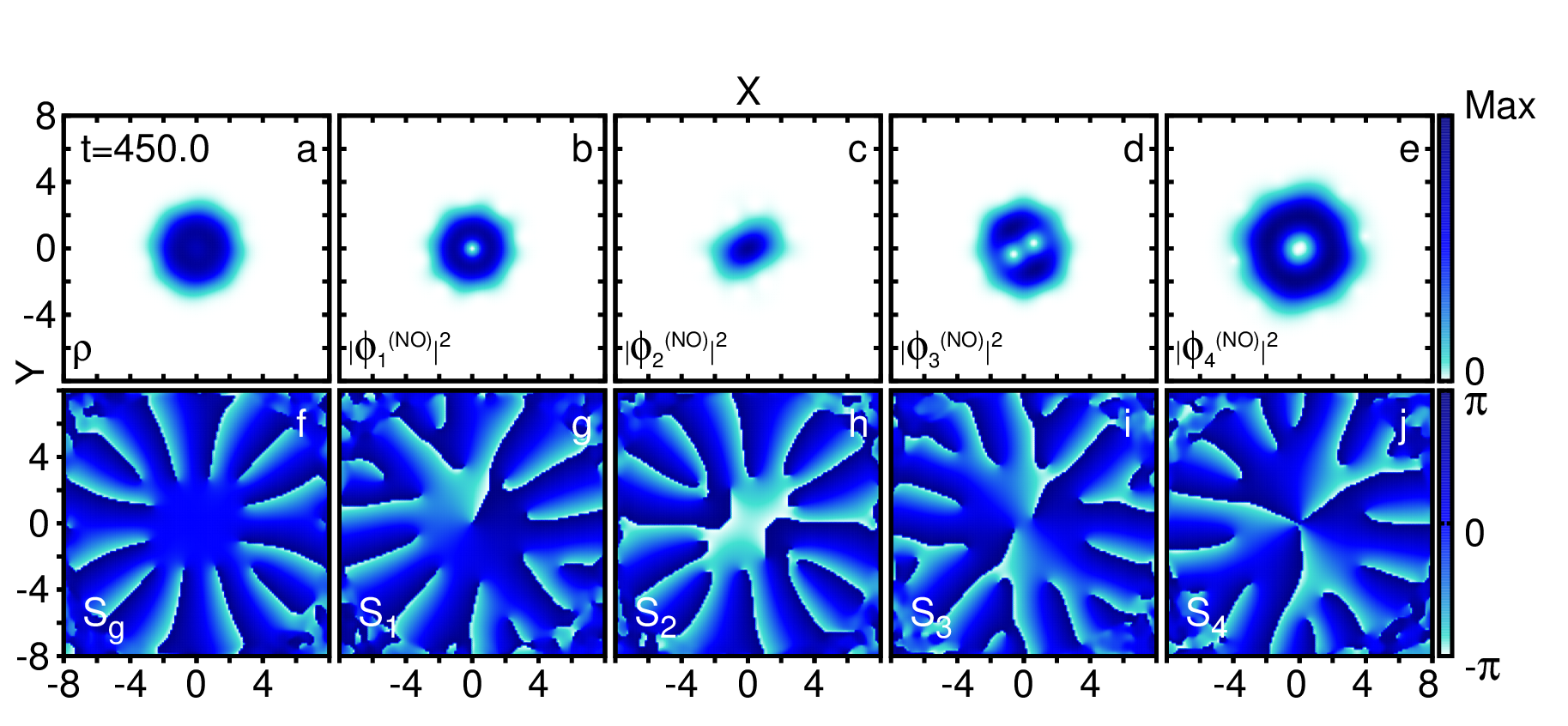}
\caption{
[COLOR]
  (a) The density $\rho(\mathbf{r})$ shows a density minimum at the origin which is $75\%$ the maximum density, but no true node.
  (b)-(e) The natural orbital densities, $\vert \phi_i (\mathbf{r})\vert ^2$ ($i=1,2,3,4$), are plotted. 
  (f) The phase, $S_g(\mathbf{r}\vert0)$, of $g^{(1)}(\mathbf{r}\vert0)$ is plotted. Note, $S_{g}$ is not the many-body phase, $S_{\text{MB}}(\mathbf{r}_1,...,\mathbf{r}_N;t)$, which is too complicated to visualize. There are singularities in $S_g$  that are associated with phantom vortices. Since $\phi_1$ and $\phi_4$ have a node at $\mathbf{r}'=0$, they do not contribute to $S_g$.
  %which do not correspond to features in $\rho$.
  (g)-(j) The phases, $S_i$, of the natural orbitals $\phi_i$ ($i=1,2,3,4$) are plotted. Each $2\pi$ phase discontinuity marks a phantom vortex core. 
  The depicted phantom vortex configuration is stable for more than $100$ trap periods. The phantom vortices in $\phi_1$ (b),(g) and $\phi_4$ (e),(j) nucleated along preexisting topological defects through node mutation whereas the pair of phantom vortices in $\phi_3$ (d),(i) was nucleated by slow orbital-orbital vortex transfer as ghost vortices which then moved to the center. The nucleation of the phantom vortex pair was coupled to the destruction of a pair in $\phi_2$ (c),(h)  See the coupled oscillation of $(L_z)_{22}$ and $(L_z)_{33}$ [Fig.~\ref{fig:Lz}c) and Video S1 (Supplementary Information)].  
  All panels are plotted at $t=450.0$, when $\rho^{(NO)}_1=40.8\%$, $\rho^{(NO)}_2=25.0\%$, $\rho^{(NO)}_3=20.8\%$, and $\rho^{(NO)}_4=13.5\%$.
  See complementary Video S1 (Supplementary Information). All quantities shown are dimensionless.}
 \label{fig:450}
 \end{figure}

We have found that there may exist many vortices in the natural orbitals despite there being no visible vortices in the density (Figs.~\ref{fig:115} and \ref{fig:450}), even with angular momentum per particle $>1\hbar$ (Fig.~\ref{fig:Lz}). We name the vortices which exist in the natural orbitals \textit{phantom vortices}. 
These phantom vortices persist on long time scales compared to the period of the harmonic trap. Phantom vortices that exist for long times near the center of the trap either nucleate on existing topological defects present in the initial natural orbitals or nucleate via a transfer of vorticity between natural orbitals (Video S1). 
Since the simulation starts with a condensed state and both mechanisms of phantom vortex nucleation rely on the dynamics and occupation of several fragments, they are inherently many-body phenomena that cannot be described by mean-field methods.

We now give a chronological description of the evolution of $N=100$ particles with $M=4$ orbitals starting in the ground state at $t=0$ to $t=500$ (see also Video S1 in Supplementary Information). 
At $t=0$, $\phi_1,...,\phi_4$ resembled 1s, 2p$_{x}$, 2p$_{y}$, and 2s orbitals, respectively.  The system began with $\rho_{1}>99.7\%$ occupation in $\phi_1$ and thus was almost entirely condensed (Fig.~\ref{fig:Lz}b).  As the anisotropy was ramped up, $\phi_1 ,..., \phi_4$ deformed smoothly and rotated with the potential.  
By $t=29.6$, due to a reordering of the natural occupations, $\phi_3$ and $\phi_4$ switched labels. At this time, $\phi_1$, $\phi_3$, and $\phi_4$ already showed faint hints of vortices at the edge of their density similar to those seen at later times in Figs.~\ref{fig:115} and \ref{fig:450}. These vortices are the fragmented counterparts of \textit{ghost vortices} reported in~\cite{Tsubota2002}.  
%Storm %
Ghost vortices are distinct from phantom vortices. Ghost vortices are phase defects outside the bulk of the condensate density. Although they do not contribute significantly to the energy or angular momentum, they may -- through interference effects -- be responsible for surface waves~\cite{Tsubota2002} which are also observed in the present work. In contrast, phantom vortices do contribute significantly to the angular momentum and energy despite being invisible in the density of the system.
 %\Storm%
By $t=45$, many ghost vortices were established in $\phi_1$ and $\phi_4$. Since the state was still more than $99\%$ condensed, the ghost vortices in $\phi_1$ manifested in the density and would thus have been detectable in high-fidelity absorption imaging.  A one-dimensional cut along a core of a ghost vortex in $\rho(\mathbf{r})$ showed that the outer density maximum was less than $1\%$ of the density maximum at the center of the cloud. By this time, the two initial lobes in $\phi_2$ had spread out and closed off the angular node into an elliptic shape. This was the first sign of a true phantom vortex in the bulk of the cloud (Fig.~\ref{fig:115}c). We term this first mechanism of phantom vortex nucleation \textit{node mutation}, since a node in the fragment deforms and mutates into a phantom vortex.
Around $t=70$, ghost vortices in $\phi_4$ fused with a distorted angular node to split the two lobes into four with an ``I'' shaped node (see Video S1 at $t=99.4$). This node mutated to nucleate three persistent phantom vortices which merged into the charge-$3$ vortex shown in Fig.~\ref{fig:450}e,j. 

By $t=90$, the system was significantly fragmented,  $\rho^{(NO)}_1 \approx 90\%$ (Fig.~\ref{fig:Lz}b). Fragmentation obscured the vortex-induced density nodes of $\phi_1$ from being visible in  $\rho(\textbf{r})$ because density from other orbitals filled the vortex cores present in $\phi_1$. We comment that although larger particle numbers delay the onset of fragmentation, we have observed phantom vortices in simulations with up to $N=10^4$ (Supplementary Information).
From $t=90$ to $t=140$, each orbital had a complicated vortex structure while the density $\rho(\textbf{r})$ was largely featureless (Figs.~\ref{fig:115}a and \ref{fig:Lz}a). This marks an important aspect of the dynamics: there were many phantom vortices in each orbital, but remarkably no vortices were detectable by directly observing the density (Figs.~\ref{fig:115} and \ref{fig:450}). In this time interval, the orbital angular momenta, $(L_z)_{ii}\equiv \langle\phi_i \vert L_z\vert\phi_i \rangle$, attained their maximal values and started to decay to their equilibrium values despite the maximal anisotropy.  

After $t=150$, the trap was still maximally anisotropic, yet the system energy and total angular momentum became saturated (Fig.~\ref{fig:Lz}d).
By $t=144$, the vortex structures of $\phi_2$ and $\phi_4$ reached their steady state. $\phi_2$ had a single phantom vortex near the center of the cloud which was nucleated from the single node of its initial shape. This node-mutated phantom vortex persisted for the length of the simulation. At $t=150$, $\phi_4$ had three phantom vortices in a linear arrangement about the center of the cloud, which persisted for the duration of the anisotropy. This triplet fused into a triply charged phantom vortex at the center of the cloud at $t=380$ when the trap became symmetric again (Fig.~\ref{fig:450}e,j). Near $t=220$, $\phi_2$ and $\phi_1$ swapped labels due to occupation reordering (see $\rho^{(NO)}_1$ and $\rho^{(NO)}_2$ in Fig.~\ref{fig:Lz}b), therefore the steady state of the phantom vortex structure can be seen in Fig.~\ref{fig:450}b,e,g,j. From $t=220$ on, the labeling of the orbitals remained fixed because the occupation numbers did not change order anymore (cf. Fig.~\ref{fig:Lz}b).

At $t=220$, there were two corotating vortices near the center of the cloud in $\phi_2$ and none present in $\phi_3$. This phantom vortex pair nucleated at the edge of the cloud and then moved towards the center, resembling mean-field vortex nucleation~\cite{Suominen2003,Tsubota2002}.  However, in our treatment, this pair was transient. The phantom vortices then transferred from $\phi_2$ to $\phi_3$, marking a second mechanism unique to phantom vortex nucleation: slow orbital-orbital vorticity transfer, which we now describe in detail.
By $t=275$, there were two prominent ghost vortices in $\phi_3$ and the phantom vortex pair in $\phi_2$ had returned from the center to the edge of this fragment.  Gradually, the vortices in $\phi_2$ disappearred by exiting the edge of the orbital while the prominent phantom vortex pair in $\phi_3$ entered the orbital bulk. By $t=340$, the phantom vortex nucleation in $\phi_3$ was complete, marked by the intervortex separation being as small as previously in $\phi_2$ at $t=220$, before the transfer (Video S1).  This direct interaction between $\phi_2$ and $\phi_3$ appears as a strong correlation in the orbital angular momentum, $(L_z)_{22}$ and $(L_z)_{33}$, as seen from $t=250$ to $t=500$ in Fig.~\ref{fig:Lz}c.  The transfer of vorticity between fragments occurred on a time scale much slower ($\tau \approx 50-100$) than the trap rotation period ($\tau\approx8$). 

As the anisotropy was ramped down, the phantom vortex pair in $\phi_3$ approached the trap center and reached a minimal separation of $\approx\hspace{-0.5mm}1.5$ by $t=380$ when the trap became symmetric. This was the only instance where we observed a phantom vortex nucleate at the edge of a fragment and persist in its bulk (Fig.~\ref{fig:450}d,i). 
%\[Storm
Simultaneously, during the ramp-down, three singly charged phantom vortices in $\phi_4$ coalesced into a dynamically stable charge-$3$ phantom vortex.  This contrasts with the known spontaneous decay of charge-$n$ vortices to $n$ single vortices ~\cite{Shin2004} and can only occur for phantom vortices for which the surrounding density of the other fragments may appartently stabilize the higher charge.
%]

\begin{figure}[H]
 \includegraphics[width=1\textwidth]{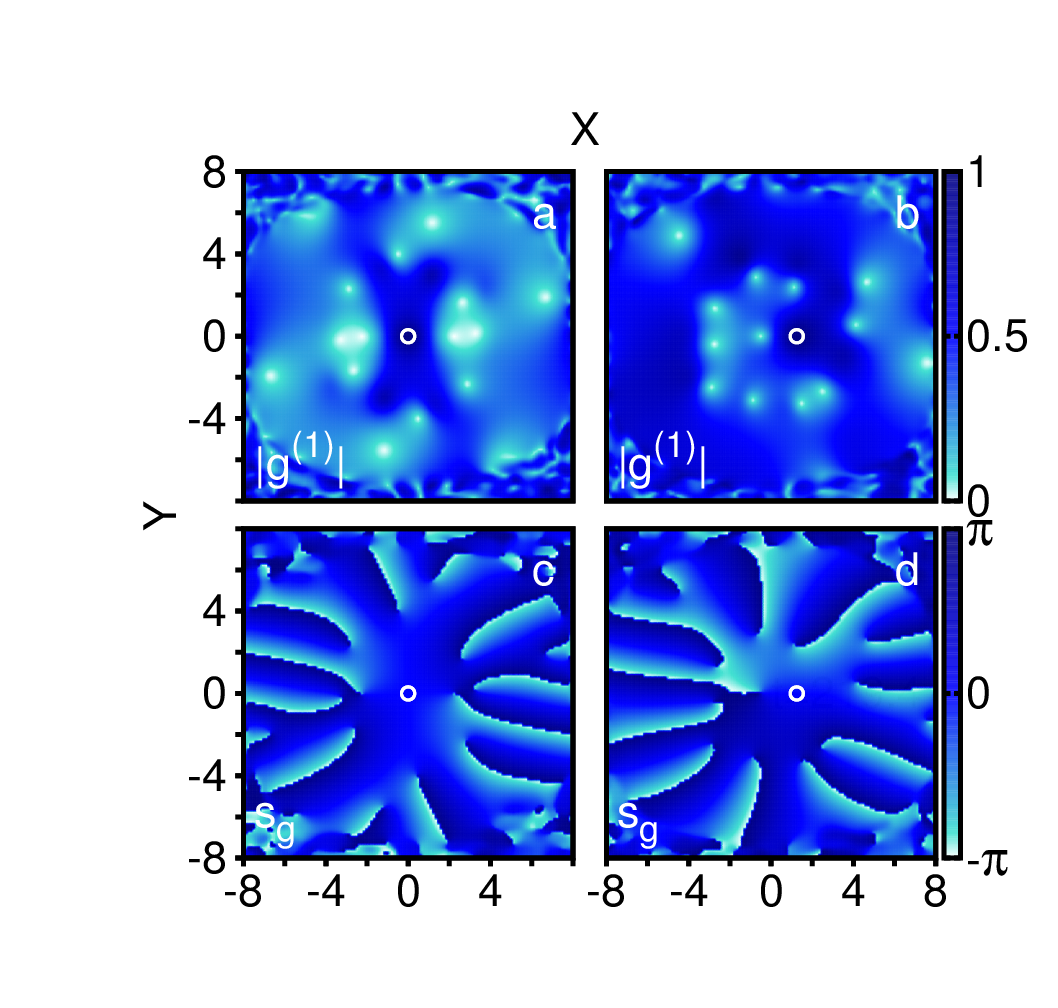}
\caption{
[COLOR]
 To visualize the four-dimensional single-particle correlation function, $g^{(1)}(\textbf{r}\vert\textbf{r}',t=219.4)$, we fix a reference point at $\textbf{r}'=(0,0)$ (a),(c) and $\textbf{r}'=(1.25,0)$ (b),(d). The function is complex, so we plot the magnitudes (a),(b) and phases (c),(d) separately.  In all panels, we mark the reference point, $\mathbf{r}'$, with a white circle.
 In panels (a) and (c) [(b) and (d)], the $\mathbf{r}'$ is colocated with a phantom vortex core in $\phi_2$ [$\phi_1$]. 
 In both cases, phantom vortex cores in all orbitals appear as spots of almost complete incoherence, $|g^{(1)}|\approx 0$, while the core colocated with the reference point has a $|g^{(1)}|\approx 1$ (full coherence). $\rho^{(NO)}_1=35.3\%$, $\rho^{(NO)}_2=35.2\%$, $\rho^{(NO)}_3=17.5\%$, and $\rho^{(NO)}_4=12.0\%$. See complementary Videos S2 and S3 for visualizations of $g^{(1)}$ at $t=115$ and $t=450$ with different reference points $\mathbf{r}'$ (Supplementary Information). All quantities shown are dimensionless.}
 \label{fig:g1}
\end{figure}

Although phantom vortices are not detectable directly from the density, they are strikingly pronounced in the one-body correlation function, $\vert g^{(1)}(\mathbf{r}\vert \mathbf{r}';t)\vert$ (Fig.~\ref{fig:g1}).  We have two key observations about the coherence of phantom vortices.  
First, we fix $\mathbf{r}'$ near the core of a phantom vortex; remarkably, $\vert g^{(1)}(\mathbf{r}\vert \mathbf{r}';t )\vert$ is close to $0$ for all $\mathbf{r}$ near and inside the cores of other phantom vortices. 
Second, we fix $\mathbf{r}'$ away from the cores of all phantom vortices and again observe that $\vert g^{(1)}(\mathbf{r}\vert \mathbf{r}' ;t)\vert$ is close to $0$ for all $\mathbf{r}$ near and inside the cores of phantom vortices (Fig.~\ref{fig:g1}, Videos S2 and S3). 
%First, we fix $\mathbf{r}'$ at the core of a phantom vortex and confirm that $\vert g^{(1)}(\mathbf{r}\vert \mathbf{r}';t)\vert  = 1 $ for $\mathbf{r}= \mathbf{r}'$.
Even phantom vortices in the same fragment are found to be incoherent.  We conclude, therefore, that phantom vortices are incoherent both with respect to each other and the remaining bulk density. This implies that they are distinct objects in the fragmented condensate. Furthermore, phantom vortices are observable since $g^{(1)}(\mathbf{r}\vert \mathbf{r}';t)$ is measurable via interference experiments, e.g. Ref.~\cite{Hofferberth2007}.

\section*{Discussion}
%[MARIOS
We have found that fragmentation increases -- together with the energy and total angular momentum of the system -- even for a rotation that is not fast enough to nucleate vortices in the density $\rho(\mathbf{r};t)$. Vortices, as seen in the laboratory (see for instance~\cite{Hodby,Fetter2009,Hall2010}), seem not emerge in our scenario. The GP mean-field description of the same system shows no vortices; however the absorbed angular momentum in that case is roughly three times less than the one predicted by the many-body theory. The angular momentum that the vortex-free GP state posseses is all due to surface excitations and deviations from a symmetric state. The many-body state, however, additionally contains angular momentum in the phantom vortices. 

It is a natural follow-up question to ask whether or not the experimentally observed vortices are coherent or fragmented  objects. If they are fragmented, phantom vortex cores across all relevant orbitals must coincide. We were able to show that if the coincident phantom vortices are of the same charge, there must be additional phantom vortices, such that the fragments can maintain orthonormality (see extended discussion in Supplementary Information). Furthermore, if all phantom vortices are coincident, they must be of different charge, see for instance $\phi_1$ and $\phi_4$ in Fig.~\ref{fig:450}b,e and Supplementary Information. In either case, and in direct contradiction to the mean-field result, the angular momentum per particle must be \textit{greater than unity} in order to nucleate a fragmented vortex in $\rho(\mathbf{r};t)$.
Perhaps some experimentally detected vortices are in fact phantom vortices rather than mean-field vortices, as suggested in~\cite{Kaspar2016}. More work is needed in that direction, in order to explore the existence and implication of fragmentation in supercritically rotated gases and the experimental visibility of phantom vortices in single-shots.
%]Marios

In summary, we have observed rich vortex dynamics within individual orbitals of a fragmented single component BEC. Since these vortex dynamics cannot be observed in the density, we termed the vortices in the fragments \textit{phantom vortices}. We have identified two mechanisms of phantom vortex nucleation, namely \textit{node mutation} and \textit{slow orbital-orbital vortex transfer}, which have no mean-field analogue. In node mutation, phantom vortices nucleate on preexisting topological defects in a fragment, whereas in slow orbital-orbital transfer, vortices are transferred between fragments. Phantom vortices are clearly visible in the correlation function. A detailed analysis of the correlation function shows that phantom vortices are completely incoherent \textit{both} with each other \textit{and} the bulk density between other phantom vortices. Phantom vortices are thus distinct quantum objects that are experimentally observable, for instance, via interference experiments.

% We suspect that phantom vortices may also play a role in the emergence of turbulence in Bose-Einstein condensates \cite{Yukalov2014}.

%may also play a role in the formation of vortices in other systems, such as the Abrikosov lattice of type-II superconductors, or the onset of turbulence in superfluid Helium.% In particular, our results suggest that there may be phantom vortex dynamics well before flux-vortices are visible in the superconductor. In turn, the existence of phantom vortices in type-II superconductors would prove the existence and importance of fragmentation and the necessity of a theoretical description which takes into account several interpenetrating order parameters in these systems.
\vspace{-0mm}
\acknowledgments{\vspace{-0.5cm}We thank Alexej I. Streltsov, Ofir E. Alon, and Tomos Wells for helpful discussions. Computation time on the Cray clusters Hermit, Hornet, and Hazel Hen at the High Performance Computing Center Stuttgart (HLRS), and financial support by the DAAD-RISE program, the Swiss SNF, the NCCR Quantum Science and Technology, DFG, and FAPESP are gratefully acknowledged. We also acknowledge the hospitality of Vanderlei Bagnato and the CEPOF at the IFSC-USP, where part of the work was completed.}

\clearpage

\section*{Author Contribution Statement}
S. Weiner ran the simulations analyzed herein, generated the plots figures and videos, and contributed to writing the manuscript. M. C. Tsatsos helped write parts of the manuscript and performed complementary calculations. A. U. J. Lode and L. S. Cederbaum conceived the idea for this project and A. U. J. Lode contributed to writing the manuscript throughout. All authors interpreted the results, reviewed and corrected the manuscript.

\section*{Additional information}
All authors declare no competing interests.

%\includepdf[pages=-]{Weiner_PhantomVortices_BEC_2016_SM_final.pdf}
\end{document}